\RequirePackage{fix-cm}
\documentclass[twocolumn,epjc3]{svjour3}  
\smartqed  % flush right qed marks, e.g. at end of proof
\RequirePackage{graphicx}
\RequirePackage{comment}
\usepackage{color}
\usepackage{lineno}
\usepackage{url}
\usepackage{amsmath}

\newcommand{\DBD}{0$\nu$DBD}

\newcommand{\CupidZ}{CUPID-0}

\newcommand{\dataset}{DataSet}
\newcommand{\ndbd}{neutrinoless double beta decay}

\newcommand{\exposure}{11.34\,kg$\times$yr} % 11.39 total, enriched + 1 natural (ch20) + DS-1000 not used for PRL analysis of 82Se
\newcommand{\limitSeventyZn}{T$_{1/2}^{0\nu\beta\beta}$($^{70}$Zn)$>$1.6$\times$10$^{21}$\,yr}
\newcommand{\limitSixtyFourZn}{T$_{1/2}^{0\nu EC \beta+}$($^{64}$Zn)$>$1.2$\times$10$^{22}$\,yr}
\newcommand{\sensitivityFourZn}{T$_{1/2}^{0\nu EC \beta+}$($^{64}$Zn)$>$1.6$\times$10$^{22}$\,yr}

\newcommand{\ckky}{\un{counts/keV/kg/y}}
\providecommand*{\un}[1]{\ensuremath{\mathrm{~#1}}}
\begin{document}
\title{Search for Neutrinoless Double Beta Decay of $^{64}$Zn and $^{70}$Zn with CUPID-0}

\author{O.~Azzolini\thanksref{Legnaro}
\and J.~W.~Beeman\thanksref{LBNL}
\and F.~Bellini\thanksref{Roma,INFNRoma}
\and M.~Beretta\thanksref{MIB,INFNMiB}
\and M.~Biassoni\thanksref{INFNMiB}
\and C.~Brofferio\thanksref{MIB,INFNMiB}
\and C.~Bucci\thanksref{LNGS}
\and S.~Capelli\thanksref{MIB,INFNMiB}
\and L.~Cardani\thanksref{INFNRoma,e1}
\and E.~Celi\thanksref{LNGS,GSSI}
\and P.~Carniti\thanksref{MIB,INFNMiB}
\and N.~Casali\thanksref{INFNRoma}
\and D.~Chiesa\thanksref{MIB,INFNMiB}
\and M.~Clemenza\thanksref{MIB,INFNMiB}
\and O.~Cremonesi\thanksref{INFNMiB}
\and A.~Cruciani\thanksref{INFNRoma}
\and A.~D'Addabbo\thanksref{LNGS,GSSI}
\and I.~Dafinei\thanksref{INFNRoma}
\and S.~Di~Domizio\thanksref{Genova,INFNGenova}
\and F.~Ferroni\thanksref{INFNRoma,GSSI}
\and L.~Gironi\thanksref{MIB,INFNMiB}
\and A.~Giuliani\thanksref{CNRS}
\and P.~Gorla\thanksref{LNGS}
\and C.~Gotti\thanksref{MIB,INFNMiB}
\and G.~Keppel\thanksref{Legnaro}
\and M.~Martinez\thanksref{Roma,INFNRoma,e2} 
\and S.~Nagorny\thanksref{LNGS,GSSI,e3} 
\and M.~Nastasi\thanksref{MIB,INFNMiB}
\and S.~Nisi\thanksref{LNGS}
\and C.~Nones\thanksref{CEA}
\and D.~Orlandi\thanksref{LNGS}
\and L.~Pagnanini\thanksref{MIB,INFNMiB}
\and M.~Pallavicini\thanksref{Genova,INFNGenova}
\and L.~Pattavina\thanksref{LNGS,GSSI,e4} 
\and M.~Pavan\thanksref{MIB,INFNMiB}
\and G.~Pessina\thanksref{INFNMiB}
\and V.~Pettinacci\thanksref{Roma,INFNRoma}
\and S.~Pirro\thanksref{LNGS}
\and S.~Pozzi\thanksref{MIB,INFNMiB}
\and E.~Previtali\thanksref{MIB,INFNMiB}
\and A.~Puiu\thanksref{LNGS,GSSI}
\and C.~Rusconi\thanksref{LNGS,USC} 
\and K.~Sch\"affner\thanksref{LNGS,GSSI,e5}
\and C.~Tomei\thanksref{INFNRoma}
\and M.~Vignati\thanksref{Roma,INFNRoma}
\and A.~Zolotarova\thanksref{CNRS} 
}

\institute{INFN - Laboratori Nazionali di Legnaro, Legnaro (Padova) I-35020 - Italy \label{Legnaro}
\and
Materials Science Division, Lawrence Berkeley National Laboratory, Berkeley, CA 94720 - USA\label{LBNL}
\and
Dipartimento di Fisica, Sapienza Universit\`{a} di Roma, Roma I-00185 - Italy \label{Roma}
\and
INFN - Sezione di Roma, Roma I-00185 - Italy\label{INFNRoma}
\and
Dipartimento di Fisica, Universit\`{a} di Milano - Bicocca, Milano I-20126 - Italy\label{MIB}
\and
INFN - Sezione di Milano - Bicocca, Milano I-20126 - Italy\label{INFNMiB}
\and
INFN - Laboratori Nazionali del Gran Sasso, Assergi (L'Aquila) I-67010 - Italy\label{LNGS}
\and
Gran Sasso Science Institute, 67100, L'Aquila - Italy\label{GSSI}
\and
Dipartimento di Fisica, Universit\`{a} di Genova, Genova I-16146 - Italy\label{Genova}
\and
INFN - Sezione di Genova, Genova I-16146 - Italy\label{INFNGenova}
\and
Universit\'e Paris-Saclay, CNRS/IN2P3, IJCLab, 91405 Orsay, France\label{CNRS}
\and
IRFU, CEA, Universit\'{e} Paris-Saclay, F-91191 Gif-sur-Yvette, France\label{CEA}
\and
Department of Physics  and Astronomy, University of South Carolina, Columbia, SC 29208 - USA\label{USC}
}

\thankstext{e1}{e-mail: laura.cardani@roma1.infn.it}
\thankstext{e2}{Present address: Fundacion ARAID and U.  Zaragoza, C/ Pedro Cerbuna 12, 50009 Zaragoza, Spain}
\thankstext{e3}{Present address: Queen's University, Kingston, K7L 3N6, Ontario, Canada}
\thankstext{e4}{Present address: Physik Department, Technische Universit\"at M\"unchen, D­85748 Garching, Germany}
\thankstext{e5}{Present address: Max-Planck-Institut f\"ur Physik, 80805, M\"unchen, Germany}


\date{Received: date / Accepted: date}
% The correct dates will be ente by the editor
\maketitle

\begin{abstract}
\CupidZ\ is the first pilot experiment of CUPID, a next-generation project searching for \ndbd. In its first scientific run, CUPID-0 operated 26 ZnSe cryogenic calorimeters coupled to light detectors in the underground Laboratori Nazionali del Gran Sasso. In this work, we analyzed a ZnSe exposure of \exposure\ to search for the \ndbd\ of $^{70}$Zn and for the neutrinoless positron-emitting electron capture of $^{64}$Zn. \\We found no evidence for these decays and set 90$\%$ credible interval limits of T$_{1/2}^{0\nu\beta\beta}$($^{70}$Zn)$>$1.6$\times$10$^{21}$ yr and T$_{1/2}^{0\nu EC \beta+}$($^{64}$Zn)$>$1.2$\times$10$^{22}$ yr, surpassing by more than one order of magnitude the previous experimental results~\cite{Belli2011}.

\end{abstract}
%\linenumbers

\section{Introduction}
\label{intro}
Double beta decay is among the rarest processes in nature. This transition, where a nucleus changes its atomic number by two units~\cite{GoeppertMayer:1935qp}, is an ideal benchmark to study atomic physics, nuclear physics as well as physics beyond the Standard Model. Despite the long half-life (10$^{18}$ to 10$^{24}$\,yr), it has been so far observed in 12 nuclei~\cite{Barabash:2015eza}.

Several extensions of the Standard Model predict that double beta decay could occur also without neutrino emission, violating the conservation of the total lepton number~\cite{Furry}. Such hypothetical transition would result in the creation of two electrons, with important implications in baryogenesis theories~\cite{Dell'Oro:2016dbc} and in particle physics, as it would naturally introduce new mass mechanisms. Finally, \ndbd\ could occur only if neutrinos and anti-neutrinos coincide, in contrast to all the other known fermions~\cite{Schechter:1982}. Thus, the observation of this transition would allow to determine the nature of this elusive particle.

The detection of \ndbd\ has been challenging the physicists community for decades. Today, lower limits of its half-life span from 10$^{24}$ to 10$^{26}$\,yr~\cite{Albert:2017owj,KamLAND-Zen:2016pfg,Alvis:2019sil,Alduino:2017ehq,Agostini:2019hzm}, and next-generation experiments are pursuing new technologies to reach a sensitivity larger than 10$^{27}$\,yr. To this purpose, future detectors will have to deploy more than 10$^{27}$ emitters (corresponding to a source mass of hundreds of kg) in background-free environments~\cite{Cremonesi:2013vla}. An energy resolution better than $\sim1$\% would also be beneficial to keep the background in the region of interest as low as possible.

Among the technologies proposed for double beta decay searches, cryogenic calorimeters stand out for their energy resolution and efficiency~\cite{Fiorini:1983yj,Pirro:2017ecr,Bellini:2018qhw}. These devices can be sketched as crystals of hundreds of grams cooled at 10\,mK and coupled to temperature sensors. Cooling the crystal at cryogenic temperature reduces its thermal capacitance $C$, so that even small energy deposits $\Delta E$ give rise to large temperature variations $\Delta T = \Delta E/C$. 
Such variations can be converted into voltage signals using dedicated temperature sensors. Those chosen by \CupidZ, namely Neutron Transmutation Doped (NTD) Ge thermistors~\cite{thermistor}, show typical voltage drops of hundreds of $\mu$V for a MeV energy deposit in the crystal.
Apart from an energy resolution better than 1$\%$, cryogenic calorimeters offer versatility in the choice of the double beta decay emitter, as the crystal can be grown from most of the isotopes of interest.

The most sensitive experiment based on the technique of cryogenic calorimeters is CUORE~\cite{Artusa:2014lgv}, that is operating a tonne-scale detector (consisting of 988 TeO$_2$ crystals) with excellent energy resolution and low background~\cite{Alduino:2017ehq,Adams:2019jhp}. 
While CUORE continues its physics programme, the CUPID collaboration (CUORE Upgrade with Particle IDentification~\cite{Artusa:2014wnl,CUPID2019}) has started to design a next-generation experiment to bring the sensitivity of cryogenic detectors above 10$^{27}$\,yr. 
The dominant source of background in CUORE are $\alpha$ particles~\cite{Alduino:2017qet}. To overcome this problem, CUPID will couple each cryogenic calorimeter to a light detector and exploit the different light yield to disentangle the $\alpha$ background from electrons~\cite{Bobin:1997qm,Pirro:2005ar}. 

This approach was developed in recent years by the LUCIFER~\cite{Beeman:2013sba,Beeman:2013vda,Beeman:2012ci,Beeman:2012jd,Beeman:2011bg,Cardani:2013mja,Cardani:2013dia,Artusa:2016maw} and LUMINEU~\cite{Barabash:2014una,Armengaud:2017hit,Buse:2018nzg,Armengaud:2015hda,Bekker:2014tfa,Armengaud:2019rll} collaborations, and gave birth to two medium-scale demonstrators: \CupidZ\ at Laboratori Nazionali del Gran Sasso, LNGS, Italy and CUPID-Mo~\cite{Armengaud:2019loe} at Laboratoire Souterrain de Modane, LSM, France.

\CupidZ\ completed its first scientific run (June 2017 -- December 2018) and was upgraded for a second scientific run, that started in June 2019. 
In this paper we present a search for the \ndbd\ of $^{70}$Zn and for the neutrinoless positron-emitting electron capture of $^{64}$Zn.

%--------------------------------------

\section{The CUPID-0 detector}
\label{detector}
The \CupidZ\ detector is an array of 26 ZnSe cylindrical crystals. Each crystal is surrounded by a plastic reflective foil (3M Vikuiti) and coupled to two light detectors, placed on its top and bottom surfaces. Most of the ``standard" light detectors do not work properly at 10\,mK. For this reason, \CupidZ\ uses small cryogenic calorimeters to convert the impinging photons into thermal signals~\cite{Beeman:2013zva}. 
These devices consist of double side polished germanium wafers (44.5\,mm diameter and 170\,$\mu$m thick) produced by UMICORE Electro-Optic Material (Geel, Belgium).
Both the ZnSe crystals and the light detectors are equipped with a NTD Ge thermistor and with a P-doped Si Joule heater. The heater injects a periodic reference pulse to enable the off-line correction for temperature variations during the data taking~\cite{Arnaboldi:2003yp,Andreotti2012}.
The detectors are disposed in five towers using a mechanical structure made of high-purity copper and PTFE elements and cooled in an Oxford $^{3}$He/$^{4}$He dilution refrigerator located in Hall-A of LNGS. The reader can find a detailed description of the \CupidZ\ design, construction, commissioning and operation in Ref.~\cite{Azzolini:2018tum}.

The main goal of the \CupidZ\ first scientific run was demonstrating the background suppression capability and understanding the residual background contributions. \CupidZ\ successfully reached these objectives, achieving the lowest background for cryogenic experiments (3.5$^{+1.0}_{-0.9}\times$10$^{-3}$\ckky\ in the region of interest for \DBD\ of $^{82}$Se at $\sim$3\,MeV) and determining its main sources~\cite{Azzolini:2019nmi}).

Besides investigating the background suppression attainable with particle identification, \CupidZ\ is the first demonstrator based on isotopically enriched crystals. Indeed, 24 of the 26 ZnSe crystals were grown starting from selenium powder 96.3$\%$ enriched in $^{82}$Se~\cite{Beeman:2015xjv,Dafinei:2017xpc}. The collaboration decided to enrich in $^{82}$Se as this is a promising emitter for double beta decay searches: it features a Q-value (2997.9$\pm$0.3 keV~\cite{Lincoln:2012fq}) well above the end-point of the natural $\beta/\gamma$ radioactivity and a relatively long half-life for the 2$\nu$DBD decay mode: 
T$^{2\nu}_{1/2}$($^{82}$Se) = ($8.60\pm0.03~(\rm{stat})^{+0.19}_{-0.13}~(\rm{syst})) \times10^{19}$~yr~\cite{Azzolini:2019yib}). 

The analysis of the data collected in the first scientific run allowed to set the most stringent limits on the half-life for the neutrinoless double beta decay of $^{82}$Se to the ground state of $^{82}$Kr (T$_{1/2}^{0\nu\beta\beta}$($^{82}$Se)$>$3.5$\times$10$^{24}$\,yr 90$\%$ credible interval~\cite{Azzolini:2019tta}) and to its 0$_1^+$, 2${_1^+}$ and 2${_2^+}$ excited states~\cite{Azzolini:2018oph}.
Moreover, the ZnSe crystals of the \CupidZ\ detector contain other two potential emitters for double beta decay: $^{64}$Zn and $^{70}$Zn. In this work we focussed on these isotopes.

%---------------------

\section{Data Production}
\label{FirstLevelAnalysis}
The data acquired by \CupidZ\ in its first scientific run are divided in ten blocks called ``\dataset". The first of them was used for the detector commissioning and was not used in the analysis of the $^{82}$Se double beta decay, as the $\alpha$ rejection tools had not yet been optimized. Given that the Q-values of the Zn isotopes lie in a region where the $\alpha$ background is negligible, we decided to include also the commissioning \dataset\ in the present analysis.
Each \dataset\ begins and ends with four days of calibration runs, performed by exposing the detector to $\gamma$ rays emitted by a $^{232}$Th source. 
We restricted our study to 22 enriched crystals plus a natural one\footnote{For this analysis we discarded three crystals of the array (two enriched in $^{82}$Se and one with natural Se) that were not showing a satisfactory bolometric performance.}, for a total ZnSe active mass of 9.18\,kg. The total collected exposure is \exposure.

The signals produced by the ZnSe crystals and light detectors were amplified and filtered with a 120 dB/decade, six-pole anti-aliasing active Bessel filter~\cite{Arnaboldi:2018yp,Carniti2016,arnaboldi20018,arnaboldi2015,arnaboldi2010,Arnaboldi:2004jj,AProgFE}. We used a custom DAQ software package to save on disk the data acquired through a 18 bit analog-to-digital  board with sampling frequency of 1~kHz for ZnSe and 2~kHz for the light detectors (which feature faster signals because of their smaller mass)~\cite{DiDomizio:2018ldc}. 
We run a derivative trigger with channel-dependent parameters on each detector to identify pulses and save a 5 (1) seconds window for pulses detected by ZnSe crystals (light detectors).
We applied a matched filter algorithm~\cite{Gatti:1986cw,Radeka:1966} to these pulses in order to suppress the most noisy frequencies, improving the evaluation of the signal amplitude. Then, we corrected the amplitudes by temperature drifts exploiting the reference pulses periodically injected by the Si resistors. The corrected-amplitudes were converted into energy using the calibration functions evaluated by attributing the nominal energy to the most intense peaks produced by the $^{232}$Th sources. Finally, we applied an algorithm that allows to improve the energy resolution of the ZnSe crystals by about 10$\%$ by removing the correlation between pulses in the ZnSe and in the corresponding light detectors~\cite{Beretta:2019bmm}.

In the last step of the data production, we searched for time-coincidences among events simultaneously triggered in more than one ZnSe crystal. This information is crucial to suppress the background for the searched signatures. To optimize the time-window in which two or more events are defined as coincident, we exposed the detector to an intense $\gamma$ source producing a sample of ``real" coincident events. This study allowed to set the optimal time-window to $\pm$20\,ms.

More details about data production techniques and algorithms can be found in Ref.~\cite{Azzolini:2018yye}.

\section{Neutrinoless Double Beta Decay of $^{70}$Zn}
\label{sec:70Zn}
$^{70}$Zn is expected to decay via 0$\nu\beta\beta$, emitting two electrons with a total energy equal to the Q-value of the transition (997.1$\pm$2.1\,keV~\cite{Audi:2017}):
\begin{equation*}
^{70}{\textnormal Zn} \rightarrow ^{70}{\textnormal Ge} + 2e^-.
\end{equation*}

Due to its poor natural isotopic abundance of (0.68$\pm$0.02)$\%$\footnote{from inductively coupled plasma mass spectroscopy}, the exposure collected for $^{70}$Zn amounts to (0.034$\pm$0.001)\,kg$\times$yr.
% or (2.95$\pm$0.08)$\times$10$^{23}$\,emitters$\times$yr.

The probability for the two electrons emitted in $\beta\beta$ decays to be fully contained in the ZnSe crystal where they are produced was evaluated through a GEANT-4 based simulation, resulting (95.67$\pm$0.46)$\%$. We searched for this process in the spectrum of events triggered in a single ZnSe crystal (``single events"), in order to suppress the background. 

\begin{figure}[thb]
\begin{centering}
\includegraphics[width=\columnwidth]{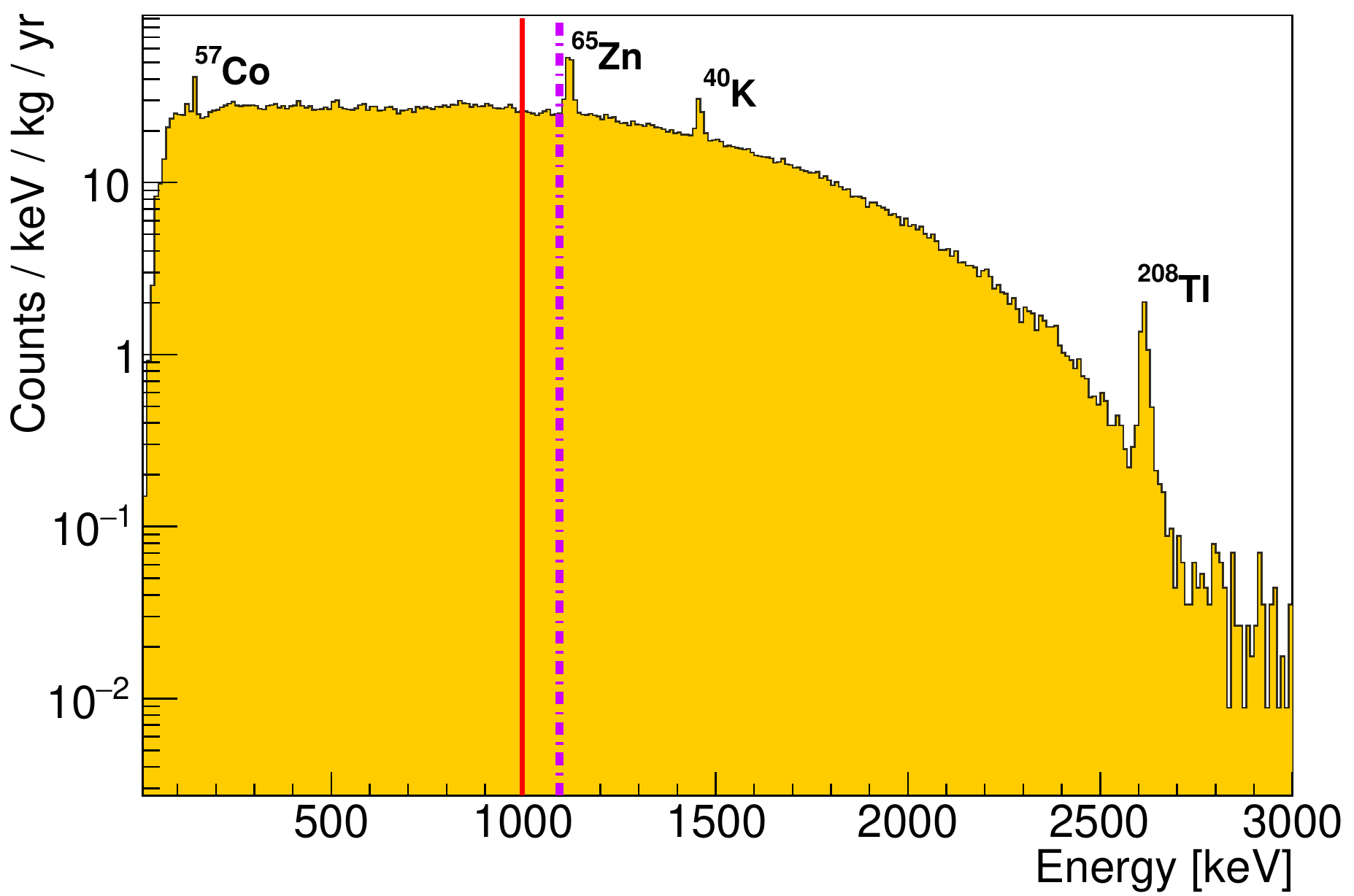}
\caption{Energy spectrum of the events detected by ZnSe crystals after data selection performed with basic cuts on the pulse shape and requiring that a single crystal in the array triggered the event. Red bar: Q-value of $^{70}$Zn (997.1$\pm$2.1\,keV). Dashed purple bar: energy of the $\gamma$ ray produced by $^{64}$Zn in Signature C (Section~\ref{sec:64Zn}). We highlight that Signature C is partly overlapped to the peak of $^{65}$Zn.}
\label{fig:SpectrumM1Zn70}
\end{centering}
\end{figure}

We selected particle-like events by applying basic cuts to the shape of the pulses recorded by ZnSe crystals. In Figure~\ref{fig:SpectrumM1Zn70} we show the energy spectrum of the single events passing these selection criteria.

The shape cuts, that allow to reject spikes due to the electronics or pile-up events, were optimised on a physical peak very close to the region of interest~\cite{Azzolini:2018yye}. For this analysis we relied on the 1115\,keV peak of $^{65}$Zn, an isotope produced via cosmogenic activation of Zn, with a relatively long half life of 244\,days. Half of the events belonging to the $^{65}$Zn peak were used to choose the values of the cuts optimising the signal-to-background ratio, while the remaining events were used to compute the  efficiency of data selection. 
The trigger efficiency and the efficiency of energy reconstruction (both $\sim$100$\%$) were evaluated using the reference pulses injected with the Si heater, following the procedure outlined in Ref.~\cite{Azzolini:2018yye}. Combining these values with the data selection efficiency computed on the $^{65}$Zn peak, we obtained a total efficiency of (95.1$\pm$0.8)$\%$. The computed efficiency was confirmed also at the energy of the $^{40}$K line at $\sim$1.46\,MeV and of the $^{208}$Tl line at $\sim$2.6\,MeV. 

We highlight that, in contrast to the analysis of $^{82}$Se 0$\nu\beta\beta$, we did not exploit the $\alpha$ rejection capability offered by scintillating bolometers, neither the aggressive time-veto described in Ref.~\cite{Azzolini:2018yye}. These analysis tools would not have been helpful, as the Q-value of $^{70}$Zn lies in a region in which the background is largely dominated by electrons produced in the 2$\nu\beta\beta$ decay of $^{82}$Se.

To compute the energy resolution at the Q-value of $^{70}$Zn, we followed the approach described in Ref.~\cite{Azzolini:2018yye}.
In cryogenic calorimeters a gaussian function is usually not able to fully describe the response to a monochromatic energy deposit~\cite{Alduino:2016zrl,Europio}. In \CupidZ\ in particular, the simplest model giving a satisfactory description of a peak consisted in the combination of two gaussian functions.
The parameters describing such functions were derived by studying the 2615\,keV peak, as this line appears in an almost background-free region. The 2615\,keV peak could be described by two gaussian functions with a mean ratio equal to 1.006, a FWHM ratio equal to 0.55 and an integral ratio equal to 0.85.
We constructed a fit model consisting of two gaussian functions with the ratios fixed to those derived using the 2615\,keV peak.
Using this model, we studied the FWHM of peaks as a function of the energy for each \dataset. We exploited the peaks produced by the $^{232}$Th source between 583 and 2615 keV and the peaks in the physics spectrum (Figure~\ref{fig:SpectrumM1Zn70}) at $\sim$145, 1120, 1460 and 2615\,keV. Due to the low trigger rate in calibration, 20 to 50 mHz depending on the crystal, the resolution did not change in runs performed with and without the calibration sources.
The dependency of the energy resolution on the energy was described using a linear function. We obtained consistent values across the ten \dataset s, excluding possible time-variations of the resolution during the CUPID-0 data-taking. 
The energy resolution extrapolated at the Q-value of the decay, averaged on the ten \dataset s, resulted $(4.45\pm0.02)$\,keV RMS.

Finally, we searched for the $^{70}$Zn 0$\nu\beta\beta$ decay signal by performing a simultaneous unbinned extended maximum likelihood (UEML) fit in a 100\,keV large analysis window centered around the Q-value.
The signal was modelled using the bi-Gaussian line shape with a mean value fixed at the position of the $^{70}$Zn Q-value. The energy resolution was fixed to the value obtained at the Q-value, and the signal decay rate $\Gamma^{0\nu\beta\beta}$ was treated as a free parameter independent from the \dataset.
We summed to this function an exponentially decreasing, \dataset-independent background, whose index was again treated as free parameter of the fit.
\begin{figure}[thb]
\begin{centering}
\includegraphics[width=\columnwidth]{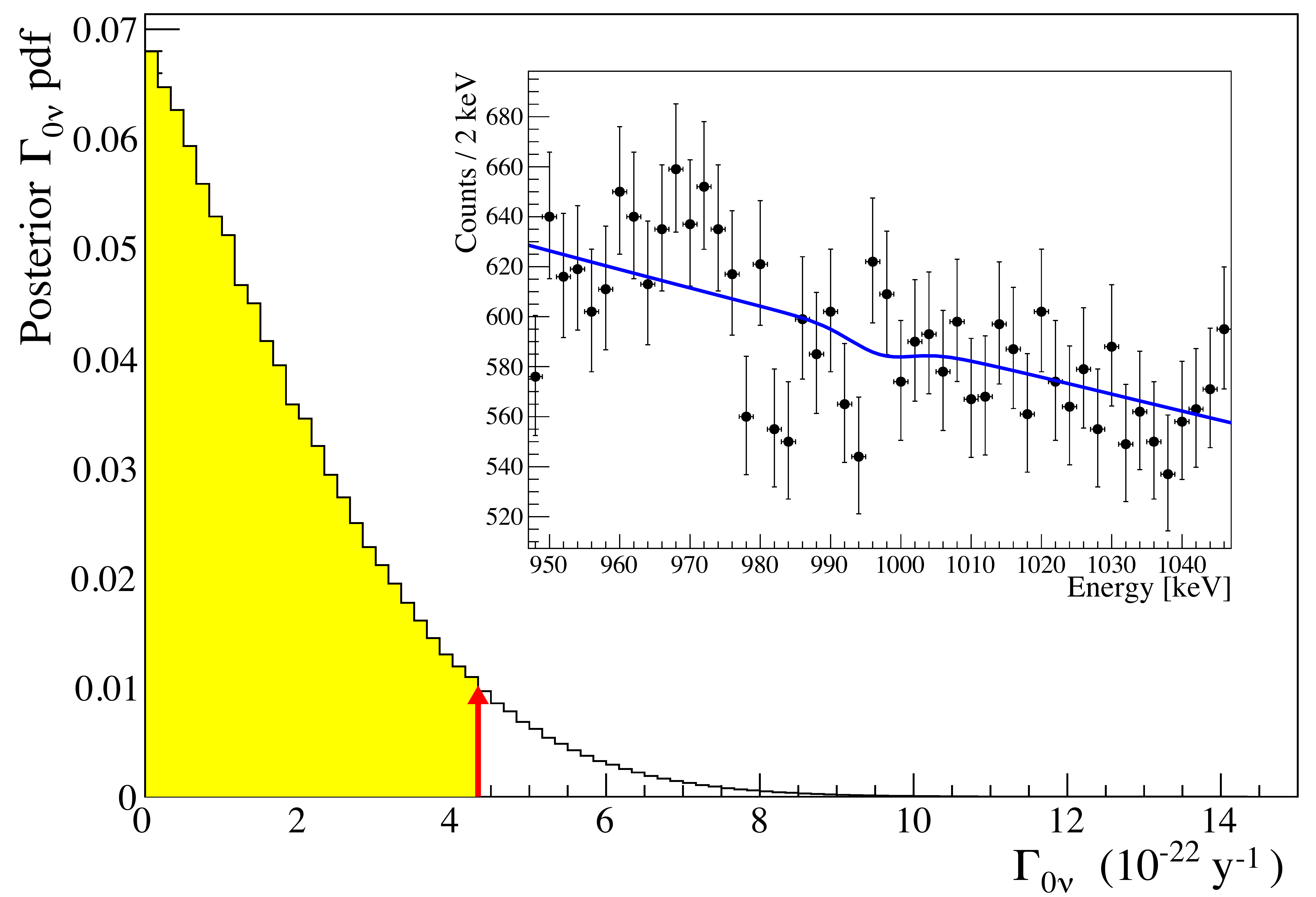}
\caption{Posterior p.d.f. of the decay rate of $^{70}$Zn. The 90$\%$ integral of the posterior is highlighted in yellow, and the red arrow indicates the value of the decay rate corresponding to the 90$\%$ credible interval. Inset: fit of the experimental data in a $\pm$50\,keV region centred around the Q-value. Dotted line: a hypothetical signal corresponding to the 90$\%$ C.I. limit set in this work.}
\label{fig:result70Zn}
\end{centering}
\end{figure}

In this study, we considered also effects due to a possible residual mis-calibration evaluated by fitting the position of the $^{65}$Zn with the same bi-gaussian model. We expect this peak to have a composite structure, as the decay of $^{65}$Zn is accompanied by the emission of X-rays or Auger electrons (8-9\,keV). If these decay products and the $\gamma$ ray are absorbed in the same crystal, we expect a peak at $\sim$1124\,keV. On the contrary, if the $\gamma$ ray is absorbed in another crystal we expect a peak at $\sim$1115\,keV. The energy resolution did not allow to resolve the two lines, so we fitted the $^{65}$Zn structure using the combination of two signal models, one for the peak at $\sim$1115\,keV and one for the peak at $\sim$1124\,keV. The position ratio of the signal models was fixed to the ratio of the nominal energies. Since the energy resolution does not vary over such a small energy range, we used the same FWHM for both the signal models. The branching ratio was a free parameter of the fit.

The mean position of the $^{65}$Zn peak was shifted by  ($-1.08\pm0.15$)\,keV with respect to its nominal value, in full agreement with the study performed in a much wider energy range exploiting the peaks produced during a $^{56}$Co calibration~\cite{Azzolini:2018yye}. This position shift was treated as a systematic source of uncertainty, independent from the \dataset. On the contrary, the energy resolution, efficiency and exposure were parameters specifically fixed for each \dataset. 
We weighted the likelihood with a Gaussian probability density function (p.d.f.) for each influence parameter, by fixing the mean and width of the p.d.f. respectively to the best-estimated values and uncertainties of each parameter.

We integrated the likelihood using a uniform non-negative prior for $\Gamma^{0\nu\beta\beta}$ and marginalizing over the background index nuisance parameter (Figure~\ref{fig:result70Zn}).
We found no evidence for the searched process in an exposure of 2.95$\times$10$^{23}$\,emitters$\times$yr and set a 90\% credible interval Bayesian lower limit on the half-life of \limitSeventyZn, surpassing by two orders of magnitude the previous limits~\cite{Belli2011,Belli2009}.
To compare this limit with the experimental sensitivity, we generated hundreds of toy Monte Carlo simulations starting from the measured background index of 26\,counts/keV/kg/yr. We fitted each simulated spectrum with a signal + background model and extracted the 90$\%$ C.I. limit from each fit. With this method we obtained a median sensitivity of T$_{1/2}^{0\nu\beta\beta}$($^{70}$Zn)$>$1.2$\times$10$^{21}$\,yr.

\section{Analysis of the $^{64}$Zn 0$\nu\beta^+$EC decay}
\label{sec:64Zn}
$^{64}$Zn features a Q-value of (1094.9$\pm$0.8)\,keV~\cite{Audi:2017} and a natural isotopic abundance of (47.5$\pm$0.1)$\%$\footnote{from  inductively coupled plasma mass spectroscopy}. This isotope can decay via electron capture with positron emission: 
\begin{equation*}
^{64}{\textnormal Zn} +  e\rightarrow ^{64}{\textnormal Ni} + {\textnormal E_{de-excitation}} + e^+
\end{equation*}
where $e$ is the captured electron, and E$_{de-excitation}$ the X-rays or Auger electrons emitted after the capture. 
Computing the containment efficiency for these de-excitation products would require a full simulation of the atomic recombination following the 0$\nu\beta^+$EC decay \cite{Bambynek}. A simpler solution is to assume every decay is followed by the emission of just one X-ray of exactly 8\,keV and to apply a volume cut corresponding to the most external layer of 27\,$\mu$m thickness of each crystal. This yield a 0.2$\%$ systematic effect on the half-life of $^{64}$Zn.

The positron emitted during the decay carries away an energy equal to (Q-value\,--\,2$m_e\sim$73\,keV). It then annihilates into two 511 keV $\gamma$'s, which can escape from the crystal giving rise to a rather complex signature. While the 73\,keV release will be always deposited in the crystal where the decay occurs, the two photons can be fully (or partly) contained in the same crystal, or they can deposit their full (partial) energy in other crystals, or totally escape detection.
The scheme of the possible signatures involving one or two ZnSe crystals is summarized in Table~\ref{Table:signatures}. Higher multiplicity events were not included in the analysis due to their low efficiency.

\begin{table}[htbp]
\centering
\caption{Possible signatures of the $^{64}$Zn electron capture - $\beta^+$ decay. In the column ``Signature", $\beta^+$ is the positron energy, while $\gamma_1$ and $\gamma_2$ are the two 511\,keV photons emitted by the positron annihilation. E$_I$ is the energy deposit in the scenario in which a single crystal is involved, while E$_I$+E$_{II}$  indicates that two crystals were hit by the decay products of  $^{64}$Zn. }
\begin{tabular}{lccc}

               & Signature 			&E$_I$  &E$_{I}$+E$_{II}$ 		  \\
               &  					&[keV]    &[keV]      \\
\hline
A	&$\beta^+$  							 &72.9 & -  \\
\hline
B	&$\beta^+$ + $\gamma_1$ 				&583.9 & -  \\
\hline	
C	&$\beta^+$ + $\gamma_1$ + $\gamma_2$ 	&1094.9 & -  \\
\hline
D	&$\beta^+$ + $\gamma_1$ + $\gamma_2$ 	&- & 72.9 + 511+ 511  \\
\hline
E	&$\beta^+$ + $\gamma_1$  				&- & 72.9 + 511  \\
\end{tabular}
\label{Table:signatures}
\end{table}

Since the analysis threshold is set at 200\,keV, we excluded from the analysis Signature A, that features a single energy deposit of 72.9\,keV.
We also discarded signature B, which would result in a peak at 583.9\,keV. At this energy, indeed, we expect a peaking background due to the 583.2\,keV $\gamma$ of $^{208}$Tl, a contaminant of the \CupidZ\ setup.
As a consequence, we restricted our analysis to the signatures C, D and E.

Signature C would result in a monochromatic peak in the spectrum of events triggered in a single ZnSe crystal  (Figure~\ref{fig:SpectrumM1Zn70}). The background index is thus simular to the one obtained in the search of the \DBD\ of $^{70}$Zn, resulting 25\,counts/keV/kg/yr. 
For this case, we followed the same procedure outlined in section~\ref{sec:70Zn} and derived the parameters of the fit at the energy of interest (Table~\ref{Table:M2}).

\begin{figure}[!h]
\begin{centering}
\includegraphics[width=\columnwidth]{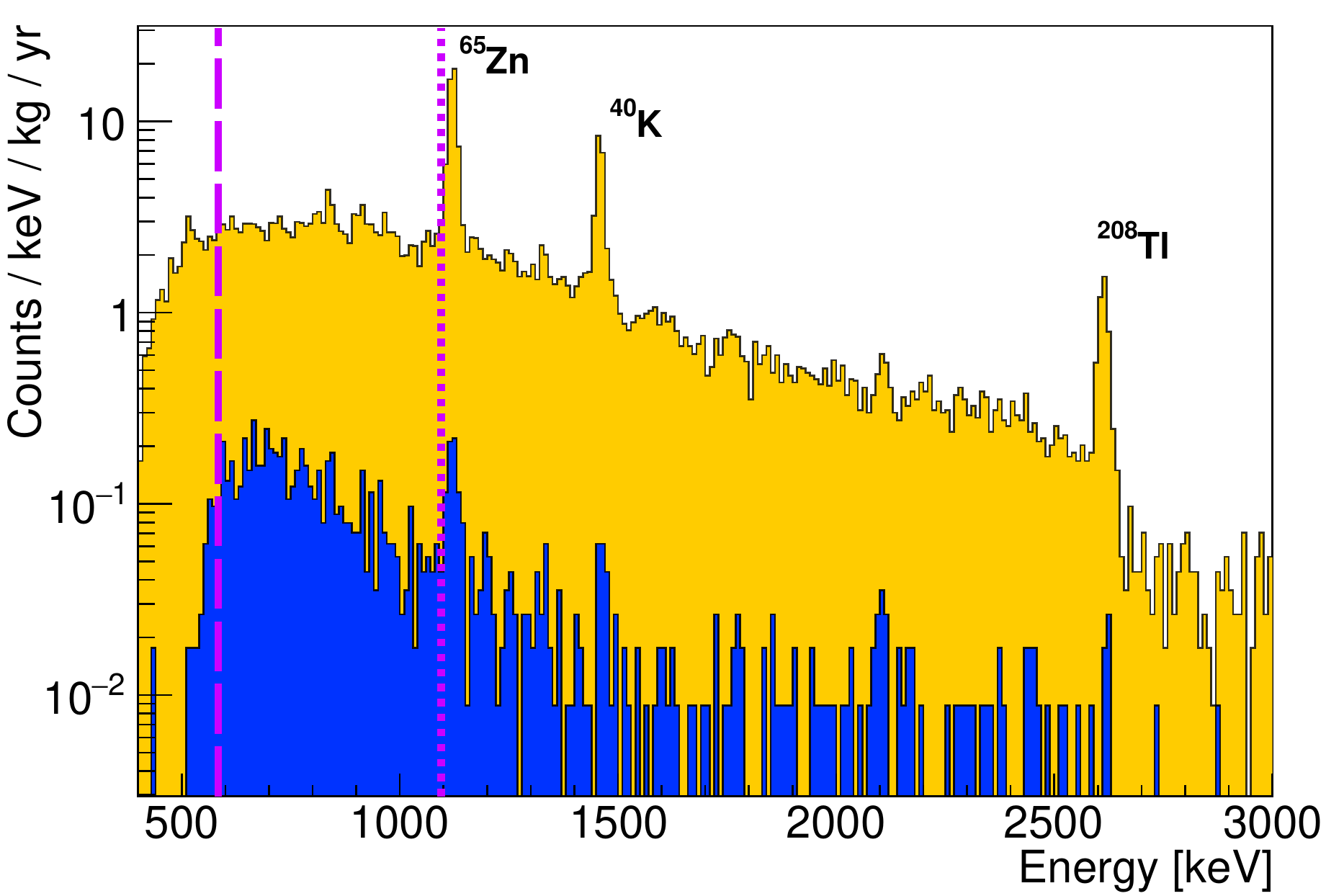}
\caption{Yellow: spectrum of the sum of the energies simultaneously released in two crystals. Blue: the same spectrum requiring that one of the two energies was equal to 511\,keV $\pm$2$\sigma$. The vertical bars indicate the total energy of the Signatures D (dotted) and E (dashed). We highlight that Signature D is partly overlapped to the peak of $^{65}$Zn.}
\label{fig:M2spectrum}
\end{centering}
\end{figure}

In Signature D, the total absorbed energy is the same as Signature C, but in this case two crystals are involved in the detection. We thus produced a spectrum by summing the energies released in two crystals (E$_{I}$+E$_{II}$), shown in Figure~\ref{fig:M2spectrum} - yellow. In this spectrum the reader can still observe the $\gamma$ peaks produced by $^{40}$K, $^{65}$Zn (which gives rise to a peaking background in the signal region) and $^{208}$Tl, while the continuum due to the 2$\nu\beta\beta$ decay of $^{82}$Se is dramatically suppressed. 

To further reduce the background in the region of interest for signature D, we required one of the two energies composing the sum spectrum to be comprised in a $\pm$2$\sigma$ interval centred around the 511\,keV peak. This cut reduces the containment efficiency from (3.07$\pm$0.06)$\%$ to (0.88$\pm$0.03)$\%$ but, at the same time, suppresses the background index from 5.6\,counts/keV/kg/yr to 4.8x10$^{-2}$ counts/keV/kg/yr, thus enhancing the sensitivity.
The spectrum obtained imposing this requirement is reported in Figure~\ref{fig:M2spectrum} - blue.

Finally, signature E should result in a peak at 72.9+511 keV in the E$_{I}$+E$_{II}$ spectrum (Figure~\ref{fig:M2spectrum} - yellow). Due to the energy threshold at 200\,keV, we could not trigger separately the 72.9\,keV and the 511\,keV energy deposits. For this reason, we did not exploit the same cut on the energy of the $\gamma$ ray adopted in the analysis of the previous signature and we obtained a background index of 2.6\,counts/keV/kg/yr. \\

The E$_{I}$+E$_{II}$ spectrum is expected to have a worse energy resolution compared to the spectrum in which the same amount of energy is released in a single crystal. 
For this reason, we repeated the study outlined in section~\ref{sec:70Zn} to determine the energy resolution at the energies of interest. 
We derived again the model describing a monochromatic energy release at 2615\,keV in the E$_{I}$+E$_{II}$ spectrum. This model was used to fit the most intense peaks produced by the $^{232}$Th source and in the physics spectrum. The peaks were the same used for the study of the spectrum in which a single crystal triggers, with the exception of the $^{57}$Co peak that, in this case, would fall below the analysis threshold. The dependency of the resolution on the energy was modelled with a linear function in the interval from 583 and 2615\,keV. This function was used to extract the energy resolution at the energy of interest for signatures D and E, reported in Table~\ref{Table:M2}.

\begin{table}[htbp]
\centering
\caption{FWHM energy resolution and containment efficiency for the three signatures of $^{64}$Zn decay.}
\begin{tabular}{lcc}

         & FWHM resolution 		&Containment efficiency		  \\
         & [keV] 				&$\%$      \\
\hline	
C	&11.37$\pm$0.09	&2.75$\pm$0.05  \\
\hline
D	&13.65$\pm$0.15	&0.88$\pm$0.03   \\
\hline
E	&9.30$\pm$0.26	&2.36$\pm$0.05   \\
\end{tabular}
\label{Table:M2}
\end{table}

In the same Table, we also report the values of the containment efficiency, derived through a Monte Carlo simulation accounting for the same analysis threshold of the experimental data.
Other contributions to the efficiency do not depend on the energy and include the trigger efficiency and the energy reconstruction efficiency. The combination of these two numbers results (98.971$^{+0.033}_{-0.034}\%$). In addition, for signature C we used the same basic cuts on the pulse shape described in section~\ref{sec:70Zn}, obtaining an event selection efficiency of (95.1$\pm$0.8) $\%$. Concerning signatures D and E on the contrary, we did not further select the events because of the lower background.

We performed a simultaneous fit to the three described spectra. The signal, as well as the peaking backgrounds such as the lines produced by the decay of $^{65}$Zn (Figure~\ref{fig:SpectrumM1Zn70} and \ref{fig:M2spectrum}), were modelled using a bi-gaussian function $\mathcal{G}$ with mean value fixed to the nominal peak position ($\mu$) and width fixed to the one derived by the resolution studies ($\sigma$, see values reported in Table~\ref{Table:M2}). We included in the fit functions also an exponential background with a number of background events ($N_{bkg}$) specific for each signature.
The number of signal events is determined by a unique decay width ($\Gamma_{^{64}Zn}$): $N_{sig}^i \propto \Gamma_{^{64}Zn} \times \epsilon_i$, where $\epsilon_i$ is the total efficiency of the searched signature.
The fitting functions can thus be written as follows:
\begin{equation}
\begin{split}
\mathcal{F}^C & = N_{sig}^C \mathcal{G}(\mu_C,\sigma_C) + N_{bkg}^C + N^C_{^{65}Zn}\mathcal{G}(\mu_{^{65}Zn},\sigma_{^{65}Zn})  \\
\mathcal{F}^D  & = N_{sig}^D \mathcal{G}(\mu_D,\sigma_D) + N_{bkg}^D + N^D_{^{65}Zn} \mathcal{G}(\mu_{^{65}Zn},\sigma_{^{65}Zn})\\
\mathcal{F}^E &= N_{sig}^E \mathcal{G}(\mu_E,\sigma_E) + N_{bkg}^E.
\end{split}
\end{equation}

\begin{figure}[!hb]
\begin{centering}
\includegraphics[width=\columnwidth]{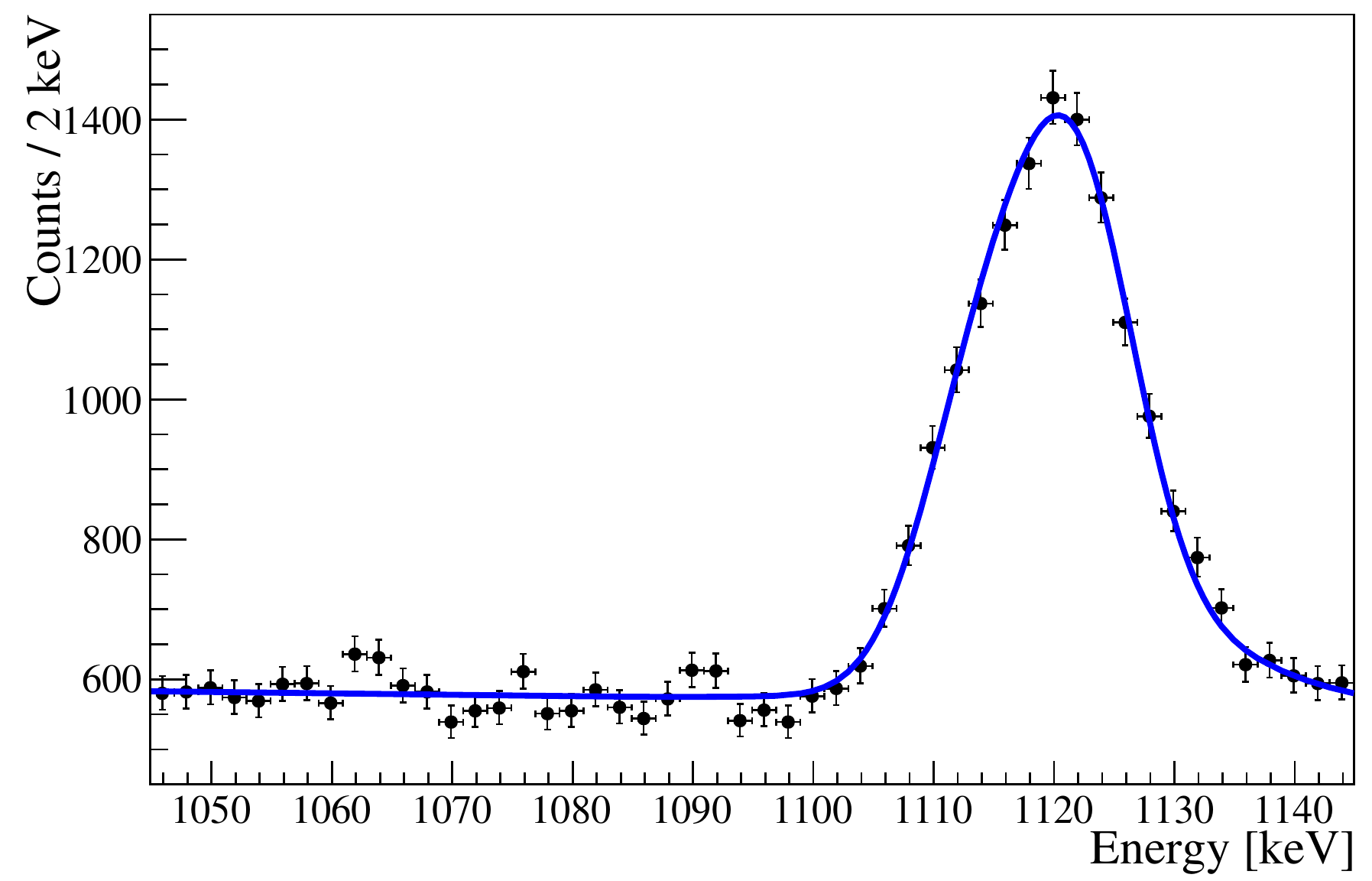}
\caption{Result of the fit of signature C. The signal is expected at E$=$1094.9\,keV. We modelled the background using an exponentially decreasing background and a peaking background due to $^{65}$Zn. Dotted line: a hypothetical signal corresponding to the 90$\%$ C.I. limit set in this work.}
\label{fig:Fit_C}
\end{centering}
\end{figure}

\begin{figure}[!hb]
\begin{centering}
\includegraphics[width=\columnwidth]{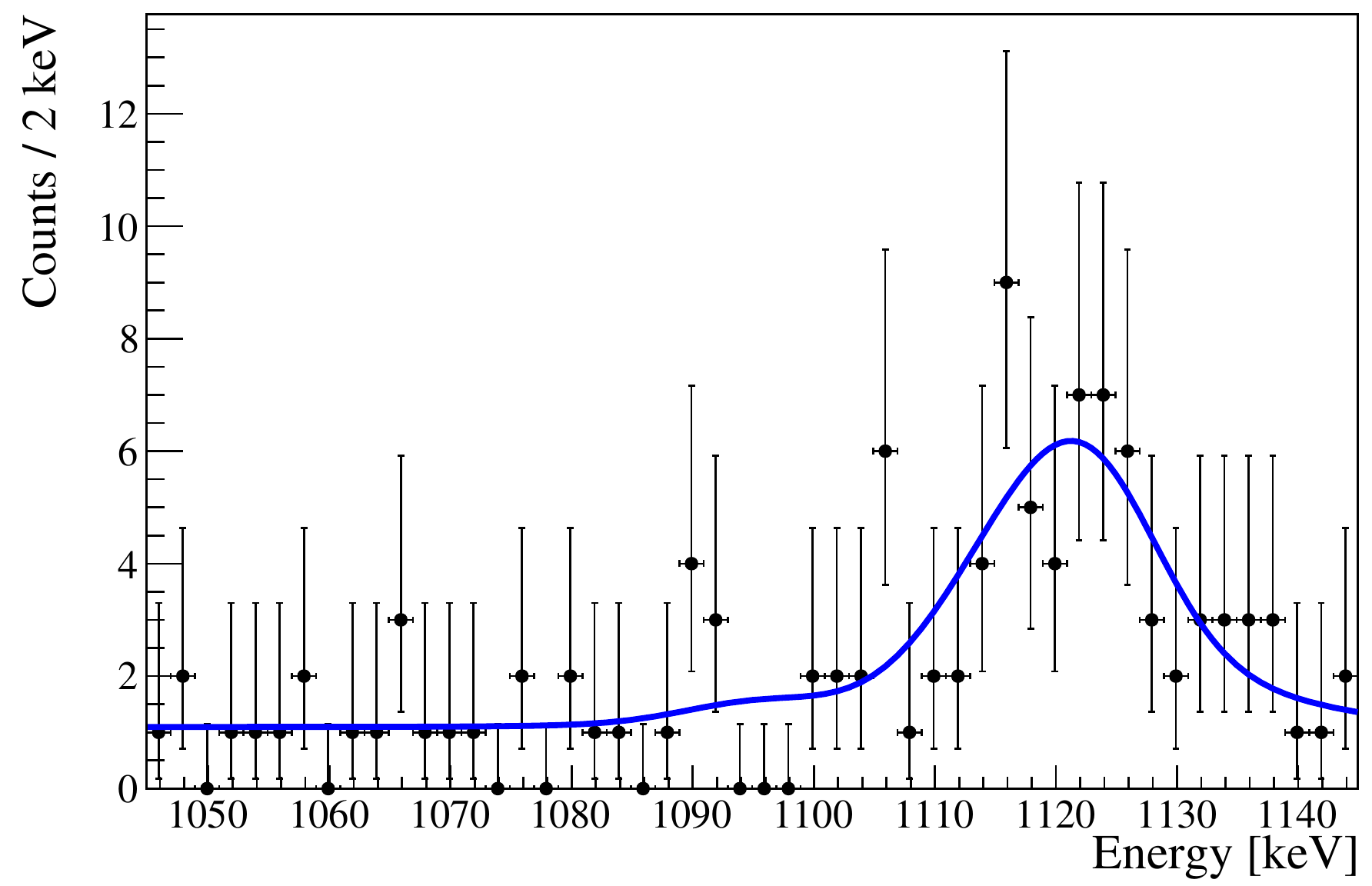}
\caption{Zoom of the blue spectrum reported in Figure~\ref{fig:M2spectrum} in the energy region of interest for signature D (1094.9\,keV). The number of events is small because we required the time-coincidence with a 511\,keV $\gamma$-ray (see main text). We fit this spectrum with the signal model, an exponentially decreasing background and a peaking background due to $^{65}$Zn. Dotted line: a hypothetical signal corresponding to the 90$\%$ C.I. limit set in this work.}
\label{fig:Fit_D}
\end{centering}
\end{figure}

\begin{figure}[!thb]
\begin{centering}
\includegraphics[width=\columnwidth]{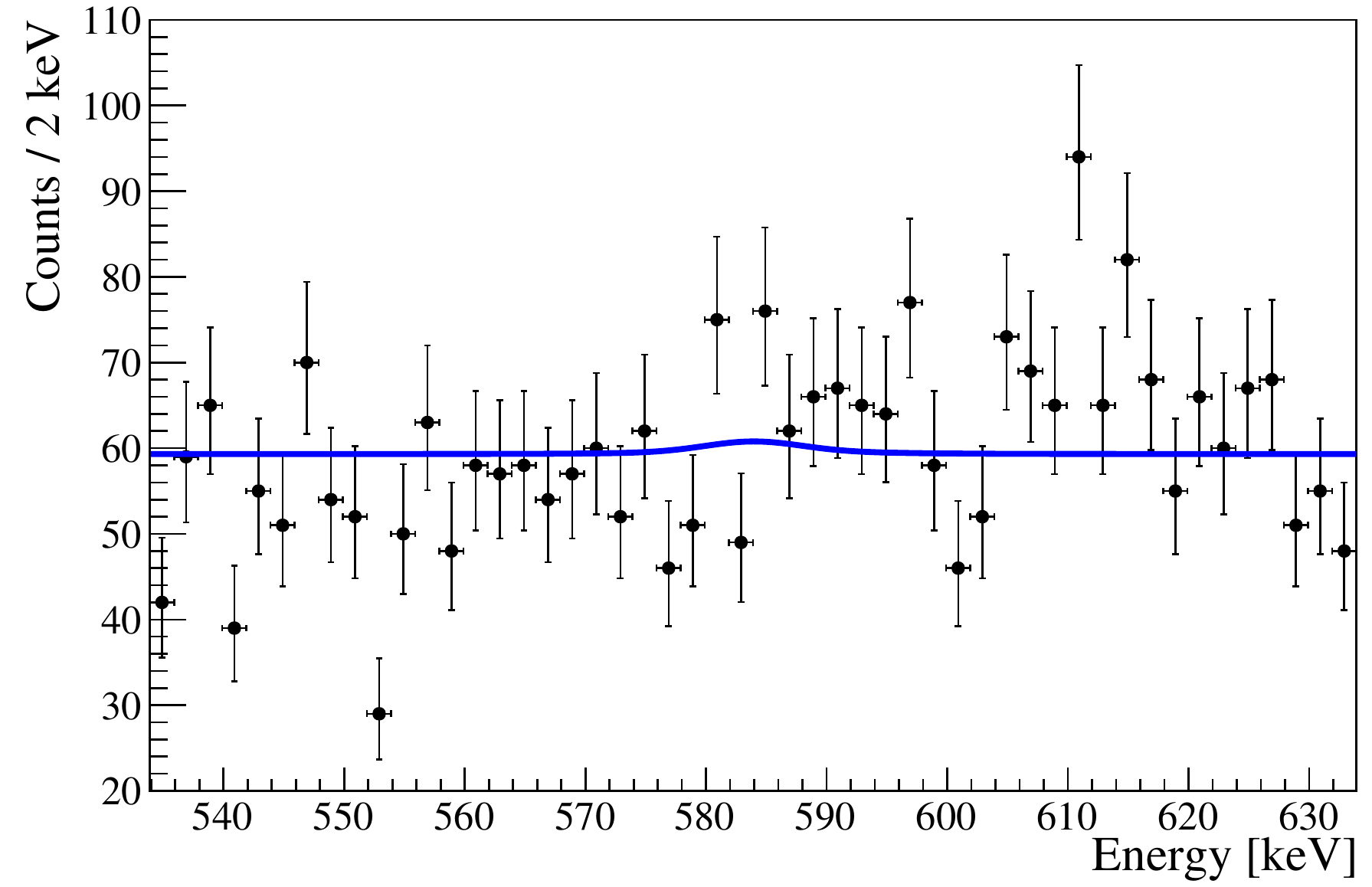}
\caption{Zoom of the yellow spectrum reported in Figure~\ref{fig:M2spectrum} in the energy region of interest for signature E (583.9\,keV). We fit this spectrum with the signal model over an exponentially decreasing background. Dotted line: a hypothetical signal corresponding to the 90$\%$ C.I. limit set in this work.}
\label{fig:Fit_E}
\end{centering}
\end{figure}

Also in this case we performed a simultaneous UEML fit. As described in section~\ref{sec:70Zn}, we included the effects of possible systematic uncertainties by weighting the likelihood with a Gaussian probability density function for each influence parameter, taking into account a possible residual mis-calibration, as well as the uncertainties on energy resolution, efficiency and exposure.
The results of the fits performed on signatures C, D and E are shown in Figures~\ref{fig:Fit_C},\ref{fig:Fit_D} and \ref{fig:Fit_E} respectively.

\begin{figure}[tb]
\begin{centering}
\includegraphics[width=\columnwidth]{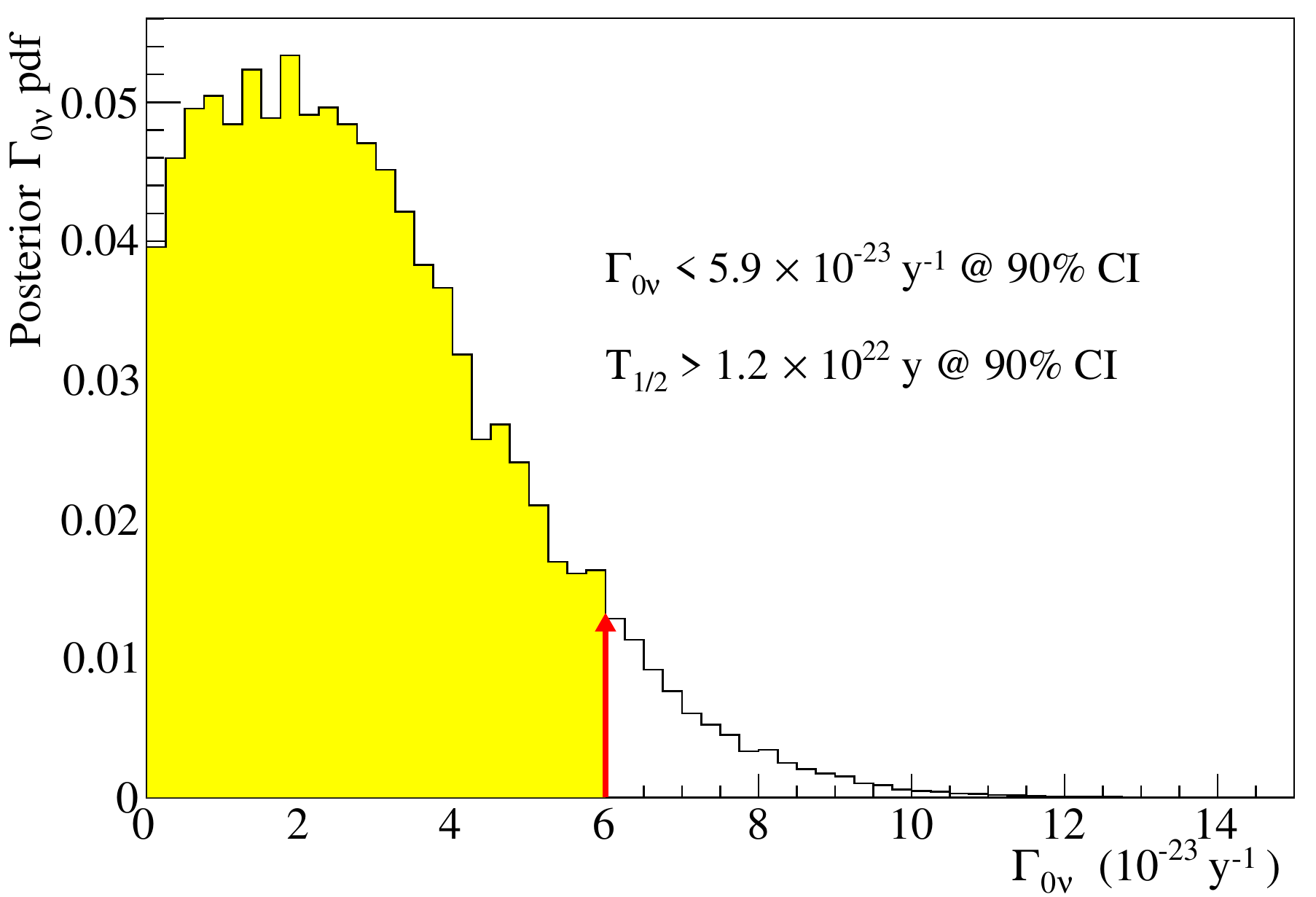}
\caption{Posterior p.d.f. of the decay rate. Yellow: integral of the posterior up to 90$\%$. The red arrow indicates the value of the decay rate corresponding to the 90$\%$ credible interval.}
\label{fig:result64Zn}
\end{centering}
\end{figure}

We chose a uniform prior for $\Gamma_{^{64}Zn}$ and integrated the likelihood marginalizing over the 
background index nuisance parameter (Figure~\ref{fig:result64Zn}).


We observed no evidence for signal and set a 90\% credible interval Bayesian lower limit on the half-life of the $^{64}$Zn electron capture - $\beta^+$ of \limitSixtyFourZn. 
This limit is slightly worse compared to the 90$\%$ C.I. median sensitivity: \sensitivityFourZn.

We underline that the obtained result largely surpasses the previous result of 8.5$\times$10$^{20}$ years reported in Ref.~\cite{Belli2011,Belli2009}, proving once more the potential of the bolometric technique.

\section{Conclusions}
In this work, we searched for the neutrinoless double beta decay of $^{70}$Zn and for the electron capture - $\beta^+$ decay of $^{64}$Zn, using the full exposure of the first \CupidZ\ scientific run of \exposure. We found no evidence of the searched processes and set lower limits on their half-life of \limitSeventyZn\ and T$_{1/2}^{0\nu EC \beta+}$($^{64}$Zn) $>$ 1.2$\times$10$^{22}$ yr, largely surpassing the previous best limits.

\begin{acknowledgements}
This work was partially supported by the Low-background Underground Cryogenic Installation For Elusive Rates (LUCIFER) experiment, funded by ERC under the European Union's Seventh Framework Programme (FP7/2007-2013)/ERC grant agreement n. 247115, funded within the ASPERA 2nd Common Call for R\&D Activities, and was funded by the Istituto Nazionale di Fisica Nucleare. We thank M.~Iannone for his help in all the stages of the detector assembly,  A.~Pelosi for constructing the assembly line, M. Guetti for the assistance in the cryogenic operations, R. Gaigher for the mechanics of the calibration system, M. Lindozzi for the cryostat monitoring system, M. Perego for his invaluable help in many tasks, the mechanical workshop of LNGS (E. Tatananni, A. Rotilio, A. Corsi, and B. Romualdi) for the continuous help in the overall set-up design. A.~S.~Zolotorova is supported by the Initiative Doctorale Interdisciplinaire 2015 project funded by the Initiatives d'excellence Paris-Saclay, ANR-11-IDEX-0003-0. 
We acknowledge the Dark Side Collaboration for the use of the low-radon clean room.
This work makes use of the DIANA data analysis and APOLLO data acquisition software which has been developed by the CUORICINO, CUORE, LUCIFER and CUPID-0 collaborations.
\end{acknowledgements}

\bibliographystyle{spphys}       % APS-like style for physics

\end{document}